# Efficient mode conversion in guiding structures with longitudinal modulation of nonlinearity

VICTOR A. VYSLOUKH,[1] YAROSLAV V. KARTASHOV,[2,*] KESTUTIS STALIUNAS[3,4]

[1]*Departamento de Fisica y Matematicas, Universidad de las Americas—Puebla, 72820 Puebla, Mexico*
[2]*Institute of Spectroscopy, Russian Academy of Sciences, Troitsk, Moscow Region, 142190, Russia*
[3]*Departament de Física i Enginyeria Nuclear, Universitat Politècnica de Catalunya, Colom 11, Terrassa 08222, Spain*
[4]*Institució Catalana de Recerca i Estudis Avançats (ICREA), Pg. Lluís Company 23, Barcelona 08010, Spain*

**We describe power-dependent dynamics of conversion of the guided modes of various guiding structures due to nearly resonant longitudinal modulation of the nonlinear coefficient of the medium. It is shown that the control of the energy exchange integrals, as well as of the input weights of the interacting modes is especially crucial for efficient mode conversion in the setting considered here. Complex dynamics of conversion incorporates various scenarios, including non-harmonic oscillations of the energy weights, which mimics Jacoby elliptical functions.**

Coupling of different modes in waveguiding optical structures is a topic of continuous interest, primarily due to numerous practical applications. In its simplest physical realization in linear optical waveguides (or fibers) with longitudinally varying refractive index, stimulated mode conversion is adequately described by a system of coupled ordinary differential equations for the modal amplitudes (see, e.g., [1] and references therein). The longitudinal modulation of the waveguide parameters modifies propagation constants of the eigenmodes, and can bring the system to phase matching if the mismatch of propagation constants of interacting modes coincides with spatial frequency of the modulation. Importantly, a simple mathematical analogy exists between stimulated mode conversion process and Rabi flopping – periodic transitions between two stationary states of a quantum system driven by the resonant external field (see [2-4] for reviews). A large variety of light-guiding architectures open many possibilities for practical realization of stimulated mode conversion: from simple one-dimensional multimode waveguides and photonic lattices [5-7], to more sophisticated helical structures, offering coupling of vortex modes with different topological charges [8-11]. More recently so called PT-symmetric waveguides, featuring spatially inhomogeneous balanced gain and absorption, were shown to influence energy conversion of eigenmodes [12]. Stimulated mode conversion is possible also in strongly guiding dielectric or metal-dielectric structures providing deep subwavelength confinement of light beams [13]. Since resonant mode coupling leads to notable modifications of the mode weights, it can be used for controllable shaping of light beams, dynamical manipulation of the diffraction strength in arrays of multimode waveguides, and transformation of the topological structure of the beam, to name just a few possibilities.

Modulation of the nonlinearity coefficient along the fiber or waveguide also leads to interesting phenomena. Thus, pulses propagating along such a fiber may develop periodically modulated envelope [14] – an effect analogous to the formation of Faraday pattern in systems with parametric excitation [15]. Such a modulation leads, for instance, to high-frequency pulsing in ring cavities built from nonlinear modulated fibers [16,17]. Interesting dynamical effects have been also obtained in fibers with dispersion modulation [18], which in some limit is equivalent to the modulation of nonlinearity coefficient. Even richer dynamics is possible when linear and nonlinear refractive index profiles in the material are simultaneously longitudinally modulated. Modern technologies allow creation of such structures, moreover, in waveguides, written in glass by femtosecond laser pulses, an increase of the linear refractive index accompanied by a decrease of the nonlinearity coefficient [19]. Similarly, fabrication of tapered photonic crystal fibers with desired longitudinal variation law of their parameters was demonstrated recently in [20]. Longitudinal nonlinearity modulation was suggested as a mechanism to control discrete solitons in waveguide arrays [21] (on [22] related idea was suggested in dissipative system). Strong longitudinal nonlinearity modulation may result in the formation of compact nonlinear excitations in waveguide arrays [23]. Recent survey of related effects can be found in [24]. Soliton dynamics was analyzed in longitudinally modulated nonlinear lattices [25,26] as well as in Bose-Einstein condensates [27]. Notice, however, that *stimulated mode conversion* due to longitudinal nonlinearity modulation was not considered so far, to the best of our knowledge.

It this Letter, we address multimode bell-shaped waveguides and more complex guiding structures in the presence of periodic longitudinal modulation of the Kerr-type nonlinearity and illustrate nontrivial features of the mode conversion process in such structures. The conversion efficiency can be high even for weak nonlinearity modulation. The key roles of the exchange integrals and the input weights of interacting modes in the transformation process are elucidated.

We consider light propagation in a multimode waveguide with longitudinal modulation of the nonlinearity coefficient, which is governed by the nonlinear Schrödinger equation for the dimensionless field amplitude $q$:

$$i\frac{\partial q}{\partial \xi} = -\frac{1}{2}\frac{\partial^2 q}{\partial \eta^2} - R(\eta)q - \mu \sin(\Omega\xi)|q|^2 q. \quad (1)$$

Here the transverse $\eta$ and longitudinal $\xi$ coordinates are scaled to the characteristic transverse width (width of the fundamental mode of the waveguide, for example) and diffraction length, respectively; the function $R(\eta)$ with $\max R(\eta) = p$ describes the transverse refractive index profile; the parameter $\mu$ stands for the depth of the longitudinal modulation of the nonlinearity coefficient, while parameter $p$ describes waveguide depth. To illustrate that mode conversion process works equally well for different waveguides, we consider two types of the structure:

sech-type waveguide $R = p\,\mathrm{sech}(\eta)$ and complex grating-type structure $R = (p/2)[1+\cos(4\eta)]\,\mathrm{sech}(2\eta/\pi)$ with multiple transverse shape oscillations. At $\mu = 0$ such waveguides support a set of stationary eigenmodes of the form $q(\eta,\xi) = w_k(\eta)\exp(ib_k\xi)$, where $b_k$ is the propagation constant and the function $w_k(\eta)$ describes the transverse mode shape. We adopt the conventional normalization for all eigenmodes: $\int_{-\infty}^{\infty} w_k^2(\eta) d\eta = 1$. Examples of symmetric linear modes $w_{1,3}$ supported by the above mentioned waveguides with $p = 3$ are shown in Fig. 1 together with refractive index distributions that are shown in the form of potential "wells", i.e. with negative sign. It is only such modes that can mutually couple due to symmetric nonlinearity modulation in Eq. (1). For the parameters of waveguides chosen above the propagation constants are given by $b_1 \approx 2.269$, $b_3 \approx 0.373$ for the sech-type potential, and by $b_1 \approx 1.315$, $b_3 \approx 0.375$ for the grating-type one. For investigation of the mode conversion we use linear superposition $q|_{\xi=0} = c_{1,\xi=0} w_1 + c_{3,\xi=0} w_3$ of symmetric guided modes at $\xi = 0$, where $c_{1,\xi=0}$, $c_{3,\xi=0}$ are the initial modal amplitudes, and consider evolution of such input in model (1). It is convenient to introduce energy weights $\nu_k = |c_k|^2$ of the modes and select input modal amplitudes in such a way that $\sum_k \nu_{k,\xi=0} = 1$ (in the absence of losses on radiation this condition holds at any distance $\xi$).

To get a preliminary insight into field dynamics, a standard approach of resonantly coupled modes [1] was considered $q(\eta,\xi) = c_1(\xi) w_1(\eta) \exp(ib_1\xi) + c_3(\xi) w_3(\eta) \exp(ib_3\xi)$, where modal amplitudes $c_k$ are functions of $\xi$ now. Substitution into Eq. (1) yields the system of ordinary differential equations for modal amplitudes:

$$\frac{dc_1}{d\xi} = i\frac{\delta\mathcal{H}}{\delta c_1^*} = +\frac{\mu}{2}[2g_{13}|c_1|^2 c_3 e^{-i\delta\Omega\xi} + g_{31}|c_3|^2 c_3 e^{-i\delta\Omega\xi} - g_{13} c_1^2 c_3^* e^{+i\delta\Omega\xi}],$$

$$\frac{dc_3}{d\xi} = i\frac{\delta\mathcal{H}}{\delta c_3^*} = -\frac{\mu}{2}[2g_{31}|c_3|^2 c_1 e^{+i\delta\Omega\xi} + g_{13}|c_1|^2 c_1 e^{+i\delta\Omega\xi} - g_{31} c_3^2 c_1^* e^{-i\delta\Omega\xi}],$$ (2)

where $\delta\Omega = b_1 - b_3 - \Omega$ is the detuning from exact resonance frequency $\Omega_r = b_1 - b_3$. Here we introduced the "exchange integrals":

$$g_{13} = \int_{-\infty}^{\infty} w_1^3 w_3 d\eta, \quad g_{31} = \int_{-\infty}^{\infty} w_3^3 w_1 d\eta,$$ (3)

and took into account the normalization for modes discussed above. The system (2) admits Hamiltonian formulation with $\mathcal{H} = (i\mu/2)(g_{13}|c_1|^2 + g_{31}|c_3|^2)(c_1 c_3^* e^{+i\delta\Omega\xi} - c_3 c_1^* e^{-i\delta\Omega\xi})$ and it conserves the sum of energy weights $\nu_1 + \nu_3$ (compare with other models of three wave coupling [28] also leading to Hamiltonian systems of coupled mode equations), Note, that in contrast with the case of linear modulation of the waveguide [5], the exchange integrals *do not involve* the shape of the waveguide $R(\eta)$ and depend solely on the shapes of the guided modes. This sets the validity limits for coupled-mode approach: the mode shape should not be strongly affected by the nonlinearity. To minimize the impact of nonlinearity on mode shapes we use in Eq. (1) the nonlinear coefficient with zero mean along $\xi$. Exchange integrals (3) have different magnitudes (typically $|g_{13}| > |g_{31}|$) and usually different signs. Thus, for sech-type potential with parameters: $g_{13} \approx 0.124$, $g_{31} \approx -0.017$, while for grating-type one: $g_{13} \approx 0.159$, $g_{31} = -0.007$, i.e. the exchange integral $g_{31}$ may be smaller than $g_{13}$ by one order of magnitude. Eqs. (2) admit implicit analytical solution at exact resonance $\delta\Omega = 0$:

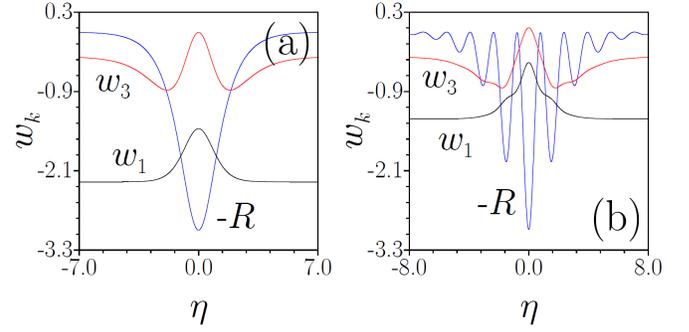

Fig. 1. Profiles of the first and third linear guided modes of (a) sech-type and (b) grating-type potentials at $p = 3$. The curves showing mode shapes $w_k$ are downshifted in accordance with corresponding propagation constants $b_k$ to illustrate the position of the corresponding energy level within potential well.

$$\frac{|(-g_{31})^{1/2} c_3 + g_{13}^{1/2}(1-c_3^2)^{1/2}|}{[\pm(g_{13}-g_{31})c_3^2 \mp g_{13}]^{1/2}} = e^{-\mu(-g_{13} g_{31})^{1/2}(\xi-\xi_0)/2},$$ (4)

where plus in the denominator should be taken when the input weight of the third mode $\nu_{3,\xi=0} > g_{13}/(g_{13}-g_{31})$, while minus corresponds to $\nu_{3,\xi=0} < g_{13}/(g_{13}-g_{31})$. In (4) $\xi_0$ is the integration constant determined by the initial conditions. The weight of the first mode follows from $c_1^2 = 1 - c_3^2$. Eq. (4) predicts that in exact resonance the evolution of mode weights is non-periodic [see Fig. 2(a) illustrating solution (4)]. At $\xi \to \infty$ both weights gradually approach constant asymptotic values $\nu_{1,\xi\to\infty} = g_{31}/(g_{31}-g_{13})$, $\nu_{3,\xi\to\infty} = g_{13}/(g_{13}-g_{31})$, independently on the initial conditions (for sech-type potential with selected parameters one has $\nu_{1,\xi\to\infty} \approx 0.122$, $\nu_{3,\xi\to\infty} \approx 0.878$). If the initial weight of the third mode is below its asymptotic value then considerable energy exchange between modes occurs – the energy first accumulates in the first mode and then gradually transfers to the third mode. In contrast, when $\nu_{3,\xi=0} > \nu_{3,\xi\to\infty}$ the energy first accumulates in the third mode and then small fraction of it returns to the first mode [see dashed lines in Fig. 2(a)], so that energy exchange is weak. The distance at which weights $\nu_{1,3}$ reach their first extrema increases with $\nu_{3,\xi=0}$ and diverges at $\nu_{3,\xi=0} = \nu_{3,\xi\to\infty}$ indicating on the qualitative modification of conversion dynamics. At $g_{31} = 0$ the solution $\nu_3 = (\xi-\xi_0)^2/[(\xi-\xi_0)^2 + 4/\mu^2 g_{13}^2]$ of Eqs. (2) becomes particularly simple (here $\xi_0$ is the corresponding integration constant).

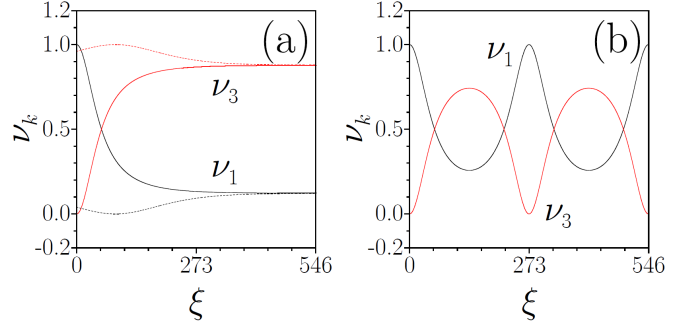

Fig. 2. Evolution of mode weights in sech-type potential calculated with the aid of coupled-mode approach for relative detuning (in percent) $\delta = 0$ (a) and $\delta = 0.1$ (b) at $\mu = 0.3$. In (a) $\nu_{3,\xi=0} = 0$ for solid lines, and $\nu_{3,\xi=0} = 0.96$ for dashed lines.

For nonzero detuning $\delta\Omega \neq 0$ the energy exchange is always periodic. Analytical solutions are not available in this case and we use numerical integration of Eqs. (2) to illustrate propagation dynamics [see Fig. 2(b)]. As discussed previously, the energy exchange between modes is strong if the initial weight of third mode $\nu_{3,\xi=0} < \nu_{3,\xi\to\infty}$ and is notably weaker for $\nu_{3,\xi=0} > \nu_{3,\xi\to\infty}$, remaining periodic in both cases [here $\nu_{3,\xi\to\infty} = g_{13}/(g_{13}-g_{31})$ is the asymptotic weight defined for exact resonance]. Apparently, the dependencies $\nu_k(\xi)$ resemble now combinations of elliptic functions. The period of oscillations diverges when $\nu_{3,\xi=0} \to \nu_{3,\xi\to\infty}$. We stress this principal difference with mode conversion process due to modulation of the linear refractive index, for which evolution of modal amplitudes is described by trigonometric functions [5]. The period of oscillations decreases with $\mu$ as follows from (2).

To describe conversion process in real-world system we now turn to integration of Eq. (1). We start with superposition of the first and third modes and calculate modal amplitudes $c_k(\xi) = \int_{-\infty}^{\infty} q(\eta,\xi) w_k(\eta) d\eta$ determined by projections of the total field amplitude at distance $\xi$ on eigenmodes of the waveguide and corresponding energy weights $\nu_k(\xi)$. We also varied the frequency detuning $\delta\Omega = \delta(b_1 - b_3)$ from exact resonance, where $\delta$ is the relative detuning.

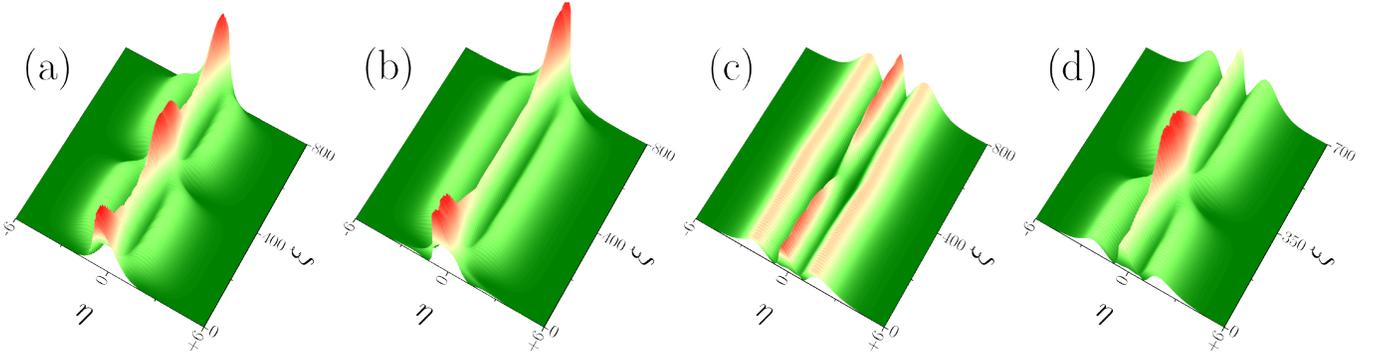

Fig. 3. Dynamics of stimulated mode conversion in sech-type potential for $\mu = 0.3$. Intensity distributions for conversion process $w_1 \to w_3$ are shown for (a) $\nu_{3,\xi=0} = 0$ and (b) $\nu_{3,\xi=0} = 0.22$, while conversion process $w_3 \to w_1$ is shown for (c) $\nu_{1,\xi=0} = 0.12$ and (d) $\nu_{1,\xi=0} = 0.13$. In all cases $\nu_{1,\xi=0} + \nu_{3,\xi=0} = 1$.

Figure 3 illustrates representative propagation dynamics ($|q(\eta,\xi)|^2$ distributions) under the condition of exact resonance ($\delta = 0$) in sech-type potential. Thus, panels (a),(b) illustrate elliptic-type oscillations of the entire distribution upon conversion process $w_1 \to w_3$, when $\nu_{3,\xi=0}$ is relatively small. Conversion length $L_3^{\max}$ at which the weight of the third mode acquires its first maximum grows with increase of $\nu_{3,\xi=0}$. Specific features of $w_3 \to w_1$ conversion dynamics encountered when input weight of the third mode $\nu_{3,\xi=0}$ is large and close to the asymptotic value $\nu_{3,\xi\to\infty}$ predicted by coupled-mode approach, are illustrated in panels (c),(d). For instance, when $\nu_{3,\xi=0} = 0.88 > \nu_{3,\xi\to\infty}$ [panel (c)] one clearly observes that at the initial stage of propagation the energy is transferred from weak first mode into strong third mode, resulting in the on-axis dip in the total intensity distribution, and in third mode domination at all propagation distances. For slightly smaller input weight of the third mode $\nu_{3,\xi=0} = 0.87 < \nu_{3,\xi\to\infty}$ [panel (d)] one observes an onset of deeply elliptic-type oscillations, with considerable power transfer between different modes. Similar types of dynamics were encountered in grating-type potential, where qualitative modifications occur now when input weight of the third mode approaches $\nu_{3,\xi\to\infty} \approx 0.957$. Similar evolution regimes encountered in two potentials considered here indicate that mode conversion due to nonlinearity modulation can be observed for any transverse trapping mechanism.

Figures 4(a) and 4(b) illustrate examples of nearly periodic evolution of mode weights for $w_1 \to w_3$ resonant conversion process in sech-type potential as predicted by Eq. (1). Weight of the second mode $\nu_2$ remains negligible and is not shown. Notice remarkable differences with non-periodic evolution from Fig. 2(a) obtained with Eq. (2). The origin of this difference comes from the fact that Eqs.(2) were derived under assumption of invariable mode profiles and they do not account for modification of $b_{1,3}$ values due to nonlinearity. In practice, the nonlinear contribution to instantaneous propagation constants results in effective frequency detuning from exact resonance. This detuning grows with the nonlinearity modulation depth and leads to finite conversion length in real-world system, which decreases with $\mu$ [see Fig.6(b) below that illustrates that conversion length diverges as $\mu \to 0$]. Radiative losses that unavoidably appear in Eq. (1) may also affect conversion dynamics.

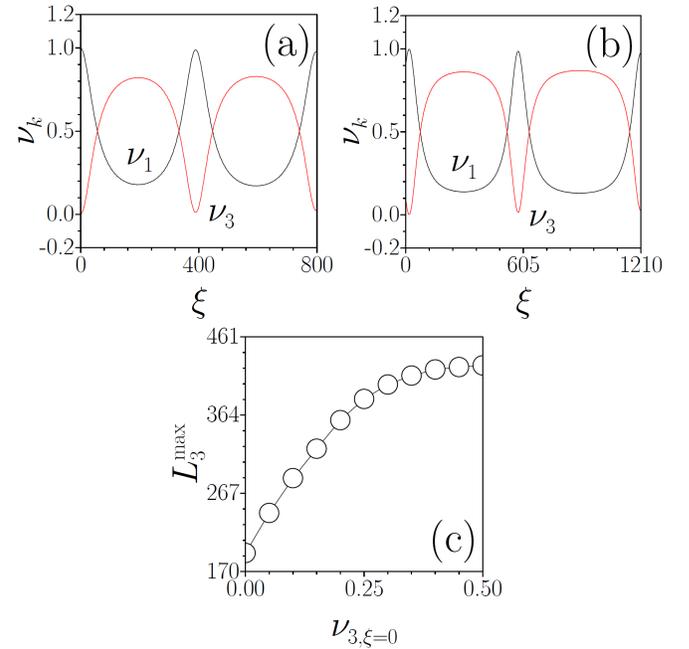

Fig. 4. Dynamics of the mode weights during $w_1 \to w_3$ conversion process calculated with Eq. (1) for different levels of seed mode $\nu_{3,\xi=0} = 0$ (a), $\nu_{3,\xi=0} = 0.1$ (b) in sech-type potential. (c) Conversion length versus $\nu_{3,\xi=0}$. In all cases $\mu = 0.3$, $\delta = 0$, and $\nu_{1,\xi=0} + \nu_{3,\xi=0} = 1$.

The presence of even weak third mode at the input drastically increases conversion length [notice different horizontal scales in (a),(b)]. At the initial stage the power flows from weak third into strong first mode (exactly as found in coupled mode approach). The conversion length $L_3^{\max}$ monotonically grows with $\nu_{3,\xi=0}$ at least up to $\nu_{3,\xi=0} = 0.5$ [Fig.4(c)]. Mode conversion due to longitudinal nonlinearity modulation is a resonant effect and the efficiency of this process strongly depends on the detuning of actual modulation frequency from the resonant value

$b_1 - b_3$. The dependence of the maximal weight $\nu_3^{\max}$ of the third mode acquired upon propagation for $w_1 \to w_3$ process is shown in Fig. 5(a) depending on detuning $\delta$. The fact that $\nu_3^{\max}$ remains below 1 even at resonance is an intrinsic property of the system with nonlinearity modulation and is not the result of radiative losses that are negligible, for example, in Fig. 4(a). The width of the resonance curve monotonically grows with the nonlinearity modulation depth $\mu$. The resonant character of the process is also evident from the dependence $L_3^{\max}(\delta)$ [Fig. 5(b)], where it is seen that conversion is slowest (but most efficient) exactly at resonance. The results agree qualitatively with predictions of coupled-mode approach (2) (red dots), but one can see that the latter provides narrower resonance curves [notice that $L_3^{\max}$ diverges at $\delta = 0$ in Eq. (2)].

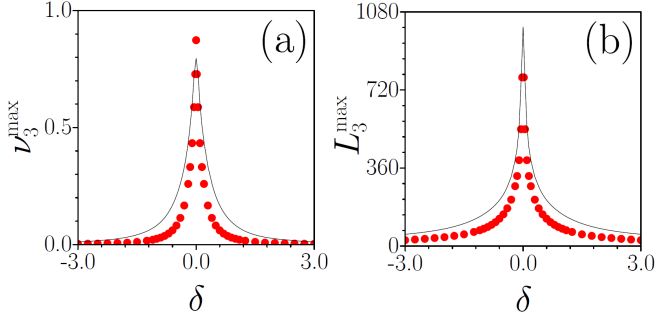

Fig. 5. Dependencies of the maximal weight of the third mode (a) and conversion length (b) on the relative detuning $\delta$ of the modulation frequency from its resonant value (in percent) in sech-type potential at $\mu = 0.05$, $\nu_{1,\xi=0} = 1$, $\nu_{3,\xi=0} = 0$. Solid lines were obtained with direct solution of Eq. (1), while red dots indicate predictions of coupled-mode approach (2).

As it was pointed out above, the efficiency of "reverse" conversion process $w_3 \to w_1$ strongly depends on the input weight of the third mode, as illustrated in Figs. 3(c) and 3(d). While for large input weights $\nu_{3,\xi=0}$ (small $\nu_{1,\xi=0}$) the conversion efficiency is strongly reduced due to the fact that nearly all power remains confined in the third mode, below critical input weight $\nu_{3,\xi\to\infty}$ one observes efficient conversion with deep oscillations in total intensity distribution. The transition between two regimes is accompanied by drastic increase of conversion length $L_1^{\max}$ (defined as a distance at which weight of the first mode acquires maximal value) at $\nu_{3,\xi=0} = \nu_{3,\xi\to\infty}$ and its subsequent rapid diminishing with decrease of $\nu_{3,\xi=0}$ [notice that in Fig. 6(a) we show $L_1^{\max}$ as a function of $\nu_{1,\xi=0}$]. One should take into account that conversion lengths $L_3^{\max}$ and $L_1^{\max}$ introduced for $w_1 \to w_3$ and $w_3 \to w_1$ processes, respectively, are *different*. This becomes evident from Fig. 4(a) if one starts from the point, where $\nu_{1,\xi} = \nu_{3,\xi} = 0.5$, and identifies the distance up to the first maxima for the first and third modes. The dependence similar to Fig. 6(a) was obtained also for grating-type potential, with the only difference that the jump in conversion length shifts to lower $\nu_{1,\xi=0} \approx 0.043$ value. Notice, that since this threshold depends on the particular exchange integrals for each type of potential, one can optimize mode conversion process by minimizing $g_{31}$ integral after proper tuning of parameters of the guiding structure.

In all cases the conversion length is a monotonically decreasing function of the nonlinearity modulation depth $\mu$ [Fig. 6(b)], that is in complete agreement with the prediction of the coupled-mode approximation. A similar reduction of the coupling length was encountered when input modes with larger power (normalization constant) $\int_{-\infty}^{\infty} w_k^2(\eta) d\eta > 1$ were used at the input.

Summarizing, we have shown that longitudinal resonant modulation of the nonlinear refractive index stimulates mode conversion in multi-mode guiding structures. Dynamics of the mode weights remarkably differs from that in the guiding structures with harmonic modulation of the linear refractive index due to the strong dependence of evolution dynamics on the input conditions and exchange integrals.

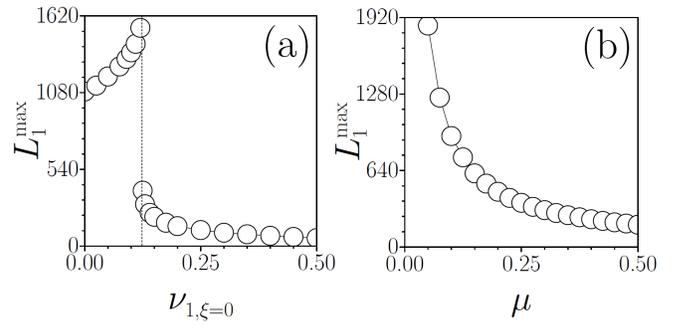

Fig. 6. (a) Discontinuous dependence of the conversion length on $\nu_{1,\xi=0}$ in sech-type potential at $\mu = 0.3$, $\delta = 0$. (b) Conversion length versus $\mu$ at $\delta = 0$, $\mu = 0.15$, $\nu_{1,\xi=0} = 0.05$ in grating-type potential. In both cases $w_3 \to w_1$ process is considered.

## References

1. D. Marcuse, Theory of Dielectric Optical Waveguides (Academic, San Diego, 1991).
2. R. W. Robinett, Phys. Rep. 392, 1 (2004).
3. I. L. Garanovich, S. Longhi, A. A. Sukhorukov, and Y. S. Kivshar, Phys. Rep. 518, 1 (2012).
4. S. Longhi, Laser and Photonics Reviews 3, 243 (2009).
5. Y. V. Kartashov, V. A. Vysloukh, and L. Torner, Phys. Rev. Lett. 99, 233903 (2007).
6. K. G. Makris, D. N. Christodoulides, O. Peleg, M. Segev, and D. Kip, Opt. Express. 16, 10309 (2008).
7. K. Shandarova, C. E. Rüter, D. Kip, K. G. Makris, D. N. Christodoulides, O. Peleg, and M. Segev, Phys. Rev. Lett. 102, 123905 (2009).
8. C. N. Alexeyev and M. A. Yavorsky, Phys. Rev. A 78, 043828 (2008).
9. C. N. Alexeyev, T. A. Fadeyeva, B. P. Lapin, and M. A. Yavorsky, Phys. Rev. A 83, 063820 (2011).
10. Y. V. Kartashov, V. A. Vysloukh, and L. Torner, Opt. Lett. 38, 3414 (2013).
11. G. K. L. Wong, M. S. Kang, H. W. Lee, F. Biancalana, C. Conti, T. Weiss, and P. St. J. Russell, Science 337, 446 (2012).
12. V. A. Vysloukh and Y. V. Kartashov, Opt. Lett. 39, 5933 (2014).
13. X. Zhang, F. Ye, Y. V. Kartashov, and X. Chen, Opt. Express. 23, 6731 (2015).
14. F. K. Abdullaev, S. A. Darmanyan, S. Bischoff, and M. P. Sørensen, J. Opt. Soc. Am. B 14, 27 (1997).
15. K. Staliunas, S. Longhi, and G. J. de Valcárcel, Phys. Rev. Lett. 89, 210406 (2002).
16. K. Staliunas, C. Hang, and V. Konotop, Phys. Rev. A 88, 023846 (2013).
17. M. Conforti, A. Mussot, A. Kudlinski and S. Trillo, Opt. Lett. 39, 4200 (2014).
18. M. Conforti, S. Trillo, A. Mussot, A. Kudlinski, Sci. Rep. 5, 9433 (2015).
19. D. Blömer, A. Szameit, F. Dreisow, T. Pertsch, S. Nolte, and A. Tünnermann, Opt. Express 14, 2151 (2006).
20. A. Bendahmane, O. Vanvincq, A. Mussot, and A. Kudlinski, Opt. Lett. 38, 3390 (2013).
21. G. Assanto, L. A. Cisneros, A. A. Minzoni, B. D. Skuse, N. F. Smyth, and A. L. Worthy, Phys. Rev. Lett. 104, 053903 (2010).
22. Y. Kominis, P. Papagiannis, and S. Droulias, Opt. Express 20, 18165 (2012).
23. F. K. Abdullaev, P. G. Kevrekidis, and M. Salerno, Phys. Rev. Lett. 105, 113901 (2010).
24. Y. V. Kartashov, B. A. Malomed, and L. Torner, Rev. Mod. Phys. 83, 247 (2011).
25. J. Zhou, C. Xue, Y. Qi, and S. Lou, Phys. Let. A 372, 4395 (2008).
26. P. Papagiannis, Y. Kominis, and K. Hizanidis, Phys. Rev. A 84, 013820 (2011).
27. D. Poletti, T. J. Alexander, E. A. Ostrovskaya, B. Li, and Y. S. Kivshar, Phys. Rev. Let. 101, 150403 (2008).
28. S. Trillo and S. Wabnitz, Opt. Lett. 16, 986 (1991).